\documentclass[a4paper]{article}
\usepackage{INTERSPEECH2021}
\usepackage{subcaption}
\usepackage{graphicx}
\usepackage{multirow}
\usepackage{graphicx}

\title{Intelligibility prediction with a pretrained noise-robust automatic speech recognition model}
\name{Zehai Tu, Ning Ma, Jon Barker}
\address{University of Sheffield, Department of Computer Science, Sheffield, UK}
\email{\{ztu3, n.ma, j.p.barker\}@sheffield.ac.uk}

\begin{document}
\maketitle

\begin{abstract}
This paper describes two intelligibility prediction systems derived from a pretrained noise-robust automatic speech recognition (ASR) model for the second Clarity Prediction Challenge (CPC2). One system is intrusive and leverages the hidden representations of the ASR model. The other system is non-intrusive and makes predictions with derived ASR uncertainty. The ASR model is only pretrained with a simulated noisy speech corpus and does not take advantage of the CPC2 data. For that reason, the intelligibility prediction systems are robust to unseen scenarios given the accurate prediction performance on the CPC2 evaluation.
\end{abstract}

\section{Introduction}
An accurate intelligibility predictor can be crucial to the development of hearing aid speech enhancement algorithms. CPC2 aims to make comparisons among speech intelligibility prediction approaches for hearing impaired listeners. 

The CPC2 database provides a large number of speech and listener recognition performance pairs. The speech signals are simulated in domestic environments and interfered by noises, music, or additional speeches. Various speech enhancement systems are used to process these speech signals and generate binaural outputs to maximise the intelligibility of the target speech for hearing impaired listeners. These processed binaural speech signals and the intelligibility of the corresponding listeners are provided in the database. The CPC2 database is divided into three partitions, each of which consists of a training set and an evaluation set. The predictions on the evaluation sets are used to measure the performance of intelligibility predictors.

The approaches in this paper follow the previous works of taking advantage of state-of-the-art ASR models for intelligibility predictions \cite{tu22_interspeech, tu22b_interspeech}. In \cite{tu22_interspeech}, an intrusive approach based on the hidden representations of an ASR model was proposed. Meanwhile, a non-intruisve approach based on ASR recognition uncertainty was proposed in \cite{tu22b_interspeech}. Unlike the approaches in the previous works, a pretrained noise-robust ASR model is used in this work, thus the training data provided in the CPC2 database is not seen by the ASR model. As a result, the proposed intelligibility predictors show a high potential to generalise to a wide range of scenarios.

\section{Method}
\label{method}

\subsection{Pretrained noise-robust ASR}
For the purpose of building generalised ASR-based intelligibility predictors, the ASR model used in this work was trained with a simulated noisy speech corpus. The Speechbrain \cite{ravanelli2021speechbrain} transformer ASR recipe for the LibriSpeech \cite{7178964}
was used for training the model, and its released parameters are used to initialise the model\footnote{https://github.com/speechbrain/speechbrain/tree/develop/recipes/\protect\\LibriSpeech/ASR/transformer} before training with the simulated noisy speech corpus.

The speech material in the simulated noisy speech corpus is LibriSpeech, which includes 960-hour read English utterances. Each utterance is convolved with a room impulse response (RIR) randomly sampled from the RIR corpus introduced in \cite{7953152}. In addition, the speech signals are interfered with by the environmental sounds in the ESC database \cite{DVN/YDEPUT_2015}. The speech-weighted signal-to-noise ratio (SNR) of each simulated noisy speech signal is randomly sampled from the range from -6 to 6\,dB. The ASR model was trained for 50 epochs and then used for intelligibility prediction.

\begin{figure}[t]
  \centering
  \includegraphics[width=0.6\linewidth]{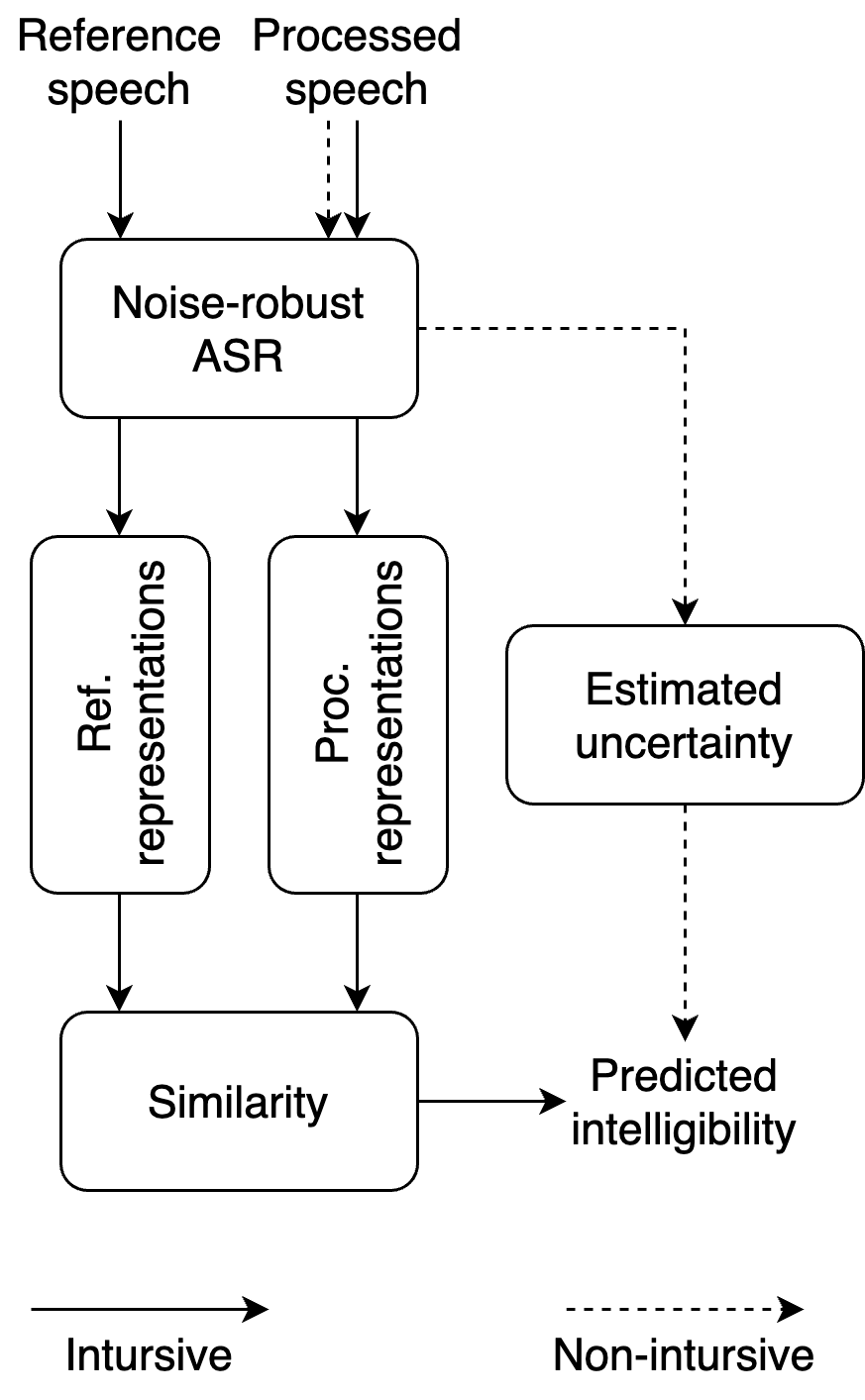}
  \caption{The work flow of the intrusive and non-intrusive intelligibility prediction methods with a noise-robust ASR model.}
  \label{fig:method}
\end{figure}

\subsection{Intelligibility prediction}
An overall work flow of both the intrusive and non-intrusive intelligibility prediction methods are shown in Figure~\ref{fig:method}. 

Given a pair of the processed speech signal and the corresponding reference signal, the intrusive method extracts the representations of the ASR hidden layer. Then the similarity between the reference and processed speech is measured and used to correlate the speech intelligibility. A detailed description of the method can be found in \cite{tu22_interspeech}. The decoder representation is used in this work, because it was shown that the language model information can help to achieve more accurate intelligibility prediction.

When the reference speech signal is not provided, the utterance-level recognition uncertainty can be estimated and then correlated to intelligibility. The detailed derivation is illustrated in \cite{tu22b_interspeech}. The negative entropy is used in this work as it takes the recognition probabilities of the tokens within the decoding beam into consideration and could produce a more accurate estimation.

The intelligibility predictors in this work do not really take hearing losses into consideration. In the first Clarity Prediction Challenge \cite{barker22_interspeech}, the evaluation results in \cite{tu22_interspeech, tu22b_interspeech} showed that the benefit by using the MSBG hearing loss simulation \cite{baer1993effects, baer1994effects, moore1993simulation, stone1999tolerable} is not significant. Also, for the purpose of building a more general intelligibility predictor, the training of the ASR model does not involve the hearing loss simulation.

Following the convention of evaluating intelligibility prediction, a logistic mapping function $f(x) = 1 / [1 + \exp(ax + b)]$ is used to re-scale the predicted intelligibility by the pre-trained ASR model. The two parameters of the mapping function were optimised with the training set.

\section{Results}
\label{results}

\begin{table}[t]
\centering
\caption{Correlation between the intelligibility by listeners and the predicted intelligibility measures on each CPC2 evaluation subset.}
\resizebox{0.6\linewidth}{!}{
\begin{tabular}{l|c|c|c}
\toprule
Subset & RMSE $\downarrow$ & NCC $\uparrow$ & KT $\uparrow$\\\midrule
\multicolumn{4}{c}{Intrusive} \\ \midrule
\textit{1} & 0.250 & 0.790 & 0.579 \\ \midrule
\textit{2} & 0.277 & 0.713 & 0.543 \\ \midrule
\textit{3} & 0.242 & 0.801 & 0.632 \\ \midrule
\multicolumn{4}{c}{Non-intrusive} \\ \midrule
\textit{1} & 0.303 & 0.660 & 0.500 \\ \midrule
\textit{2} & 0.274 & 0.715 & 0.531 \\ \midrule
\textit{3} & 0.256 & 0.773 & 0.607 \\ \bottomrule
\end{tabular}
}
\label{tab:results}
\end{table}

The evaluation results of the intrusive and non-intrusive methods on the three evaluation subsets are shown in Table~\ref{tab:results}. The root mean square error (RMSE), normalised cross-correlation (NCC), and the Kendall's Tau (KT) coefficient are reported.

\bibliographystyle{IEEEtran}
\bibliography{mybib}
\end{document}